\documentclass[preprint,showpacs,preprintnumbers,amsmath,amssymb,floatfix,endfloats*]{revtex4}

\usepackage{graphicx}
\usepackage{dcolumn}
\usepackage{bm}
\usepackage{amsfonts}

\DeclareMathAlphabet{\mathsfsl}{OT1}{cmr}{bx}{it}
\begin{document}
\title{The relationship between induced fluid structure and boundary slip in nanoscale polymer films}
\author{Nikolai~V.~Priezjev}
\affiliation{Department of Mechanical Engineering, Michigan State
University, East Lansing, Michigan 48824}
\date{\today}
%
\begin{abstract}

The molecular mechanism of slip at the interface between polymer
melts and weakly attractive smooth surfaces is investigated using
molecular dynamics simulations. In agreement with our previous
studies on slip flow of shear-thinning fluids, it is shown that the
slip length passes through a local minimum at low shear rates and
then increases rapidly at higher shear rates.  We found that at
sufficiently high shear rates, the slip flow over atomically flat
crystalline surfaces is anisotropic. It is demonstrated numerically
that the friction coefficient at the liquid-solid interface (the
ratio of viscosity and slip length) undergoes a transition from a
constant value to the power-law decay as a function of the slip
velocity. The characteristic velocity of the transition correlates
well with the diffusion velocity of fluid monomers in the first
fluid layer near the solid wall at equilibrium.  We also show that
in the linear regime, the friction coefficient is well described by
a function of a single variable, which is a product of the magnitude
of surface-induced peak in the structure factor and the contact
density of the adjacent fluid layer.  The universal relationship
between the friction coefficient and induced fluid structure holds
for a number of material parameters of the interface: fluid density,
chain length, wall-fluid interaction energy, wall density, lattice
type and orientation, thermal or solid walls.

\end{abstract}



\maketitle

\section{Introduction}

The study of fluid transport through micro- and nanochannels is
important for biotechnological applications and energy conversion
processes~\cite{Eijkel09}.  The precise control and manipulation of
fluids in systems with large surface-to-volume ratios, however,
require fundamental understanding of flow boundary conditions. Fluid
velocity profiles can be significantly modified in the presence of
slip at a solid surface.  The degree of slip is quantified by the
slip length, which is defined as a distance between locations of the
real interface and imaginary plane where the extrapolated tangential
velocity component vanishes.  It was shown by numerous experimental
studies that the main factors affecting slippage at the liquid-solid
interface include surface
roughness~\cite{Granick02,Archer03,Leger06}, surface
wettability~\cite{Churaev84,Charlaix01,SchmatkoPRL05}, fluid
structure~\cite{SchmatkoPRL05,BruceLaw09,Jacobs09}, and shear
rate~\cite{Granick01,Breuer03,Ulmanella08}.  The slip length in the
micron range is reported for complex flows near superhydrophobic
surfaces~\cite{Rothstein10} and flows of high molecular weight
polymers~\cite{JacobsRev}, while the magnitude of the slip length in
the range of a few tens of nanometers is typically measured for
flows of water over smooth nonwetting surfaces~\cite{Charlaix10}.

A number of molecular dynamics (MD) simulation studies have been
carried out to investigate the influence of structural properties of
the interface between  monatomic fluids and flat crystalline walls
on the degree of
slip~\cite{Fischer89,KB89,Thompson90,Barrat94,Nature97,Barrat99,Barrat99fd,Attard04,Priezjev07,PriezjevJCP,Li09,Asproulis10}.
It was shown that the slip length is directly related to the
intensity of structure induced in the first fluid layer by the
periodic potential from the solid substrate~\cite{Thompson90}. The
slip is reduced with increasing wall-fluid interaction energy and
due to the formation of commensurate structures between solid wall
and adjacent fluid layer. The solid walls are usually modeled as an
array of atoms arranged on sites of a periodic lattice. Two types of
walls are considered, solid and thermal, where the wall atoms are
either fixed at the lattice sites or allowed to oscillate in the
harmonic potential. It was found that the slip length weakly depends
on the value of the spring stiffness coefficient for sufficiently
strong harmonic bonds~\cite{PriezjevJCP,Asproulis10}. In addition,
the slope of the shear rate dependence of the slip length is not
significantly affected by stiff springs~\cite{PriezjevJCP}, except
at very high shear rates~\cite{LichPRL08}.

At the interface between \textit{simple fluids} and atomically
smooth, weakly attractive surfaces, the slip length is constant at
low shear rates and increases nonlinearly at higher shear rates, as
originally discovered by Thompson and Troian~\cite{Nature97} and
later confirmed by several
studies~\cite{Fang05,Priezjev07,Niavarani10,Asproulis10}. For
sufficiently strong wall-fluid interactions and incommensurate
structures of the liquid and solid phases at the interface, the slip
length varies almost linearly with shear
rate~\cite{Priezjev07,PriezjevJCP}. It should be noted, however,
that if the slip length at low shear rates is less than about a
molecular diameter then the boundary conditions for dense monatomic
fluids are rate-independent~\cite{Thompson90,Nature97,PriezjevJCP}.
Also, it was shown that molecular-scale surface roughness reduces
the magnitude of the slip length and the slope of its rate
dependence~\cite{Barrat94,Priezjev06,PriezjevJCP,Sofos09,Niavarani10}.

It was recently demonstrated that the effective slip length for
flows over anisotropic surfaces with two-component texture of
different wettability is largest (smallest) for parallel
(perpendicular) orientation of stripes with respect to the mean
flow~\cite{Vinograd09}. These conclusions hold when the stripe width
is comparable to the molecular diameter~\cite{Priezjev05}. For the
transverse orientation of the flow relative to the stripes, the slip
is reduced because of the molecular scale corrugation of the
composed surface potential, while for the parallel orientation, the
fluid molecules are transported along homogeneous stripes with
either no-slip or partial slip conditions, and, therefore, the
effective slip length is enhanced~\cite{Priezjev05}.  More recently,
it was observed that the slip length also depends on the crystal
lattice plane in contact with the fluid and on the lattice
orientation with respect to the flow direction~\cite{Tzeng07}. In
the present study, we will show that at sufficiently high shear
rates, the slip flow is anisotropic for atomically flat crystalline
surfaces; and, in particular, the slip length is enhanced when the
shear flow is oriented along the crystallographic axis of the wall
lattice.

Recent studies of friction between adsorbed monolayers and smooth
crystalline surfaces are relevant to the analysis of flow boundary
conditions~\cite{Robbins96,Tomassone97}. It was found that the slip
time, which represents the transfer of momentum between the adsorbed
monolayer and the substrate, is proportional to the phonon lifetime
divided by the normalized peak value of the structure factor
computed in the monolayer at the main reciprocal lattice
vector~\cite{Robbins96}. Also, the simulation results have shown
that the slip time is independent of the sliding direction if the
slip velocity of the monolayer is much smaller than the speed of
sound~\cite{Robbins96}. In the linear regime between friction force
and sliding velocity and in the range of film coverages from
submonolayer to bilayer, the slip times were computed directly from
the decay of the film velocity and from the decay of the velocity
correlation function at equilibrium~\cite{Tomassone97}.

During the last two decades, several MD studies have examined slip
boundary conditions at the interface between \textit{polymeric
fluids} and flat crystalline
surfaces~\cite{Thompson95,Manias96,dePablo96,StevensJCP97,Koike98,Tanner99,Priezjev04,Priezjev08,Servantie08,Priezjev09,Hendy09}.
The velocity profiles with stick boundary conditions were observed
when a highly viscous interfacial layer was formed because of the
strong wall-fluid interaction
energy~\cite{Tanner99,Manias96,Servantie08}, high fluid density and
pressure~\cite{Thompson95,Manias96,Priezjev09}, or chemical
structure of chain molecules~\cite{Denniston10}. The variation of
the slip length as a function of shear rate was reported for flat
polymer-solid interfaces with weak wall-fluid
interactions~\cite{Tanner99,Priezjev04,Niavarani08,Priezjev08,LichPRL08,Priezjev09,Hendy09}.
In our previous studies~\cite{Niavarani08,Priezjev08}, it was shown
that the rate dependence of the slip length acquires a local minimum
at low shear rates followed by a rapid growth at higher shear rates.
Furthermore, it was found that in a wide range of fluid densities,
the friction coefficient at the liquid-solid interface undergoes a
gradual transition from a constant value to the power-law decay as a
function of the slip velocity~\cite{Priezjev08,Priezjev09}.
Remarkably, the data for the friction coefficient at different fluid
densities and shear rates were found to collapse onto a master curve
when plotted as a function of the product of the main peak in the
structure factor and the contact density of the first fluid
layer~\cite{Priezjev08,Priezjev09}. Although promising results were
obtained, the simulations were performed at a single wall density
and only for one orientation of the fcc lattice with respect to the
shear flow direction~\cite{Priezjev08,Priezjev09}.

In this paper, molecular dynamics simulations were conducted for
twenty systems that include monatomic and polymeric fluids confined
by flat crystalline surfaces. In agreement with previous studies, we
report the nonlinear rate dependence of the slip length for systems
with weak wall-fluid interaction energies and relatively dense
walls. The simulation results indicate that the friction coefficient
(the ratio of fluid viscosity and slip length) in the
linear-response regime is a function of a single variable that is a
product of the height of the normalized main peak in the structure
factor and the contact density of the first fluid layer near the
solid wall. We will show that the onset of the nonlinear regime
between the wall shear stress and slip velocity is determined by the
diffusion of fluid monomers within the first layer.

The rest of the paper proceeds as follows. The details of molecular
dynamics simulations and parameter values for twenty systems are
described in the next section. The results for the rate-dependent
slip length, friction coefficient, and fluid structure are presented
in Section~\ref{sec:Results}. The conclusions are given in the last
section.

\section{Molecular dynamics simulation model and parameter values}
\label{sec:Model}

\begin{figure}[t]
\vspace*{-3mm}
\includegraphics[width=8.0cm,angle=0]{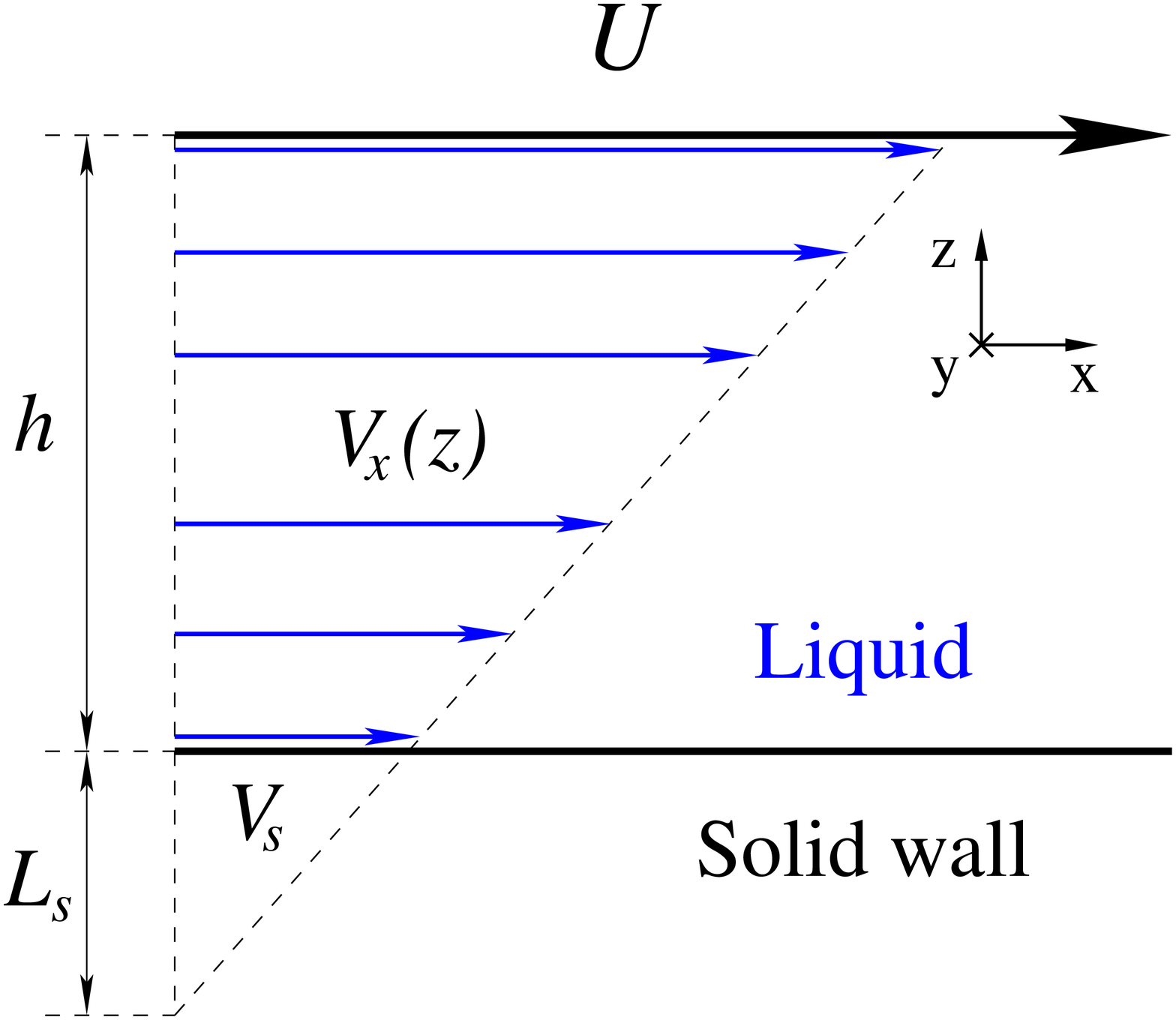}
\caption{(Color online) A schematic of the flow with slip boundary
conditions at the lower and upper walls. Shear flow is induced by
the upper wall moving with a constant speed $U$ in the $\hat{x}$
direction. The slip velocity and slip length $L_s$ are related via
$V_s\,{=}\,\,\dot{\gamma}L_s$, where $\dot{\gamma}$ is the shear
rate computed from the slope of the velocity profile.}
\label{schematic}
\end{figure}

The geometry of the computational domain and the steady flow profile
are shown schematically in Figure\,\ref{schematic}. The fluid
undergoes planar shear flow between two atomically flat walls. The
fluid phase consists of $N_{f}\!=9600$ monomers. The interaction
between any two fluid monomers is modeled via the truncated
Lennard-Jones (LJ) potential
\begin{equation}
V_{LJ}(r)\!=4\,\varepsilon\,\Big[\Big(\frac{\sigma}{r}\Big)^{12}\!-\Big(\frac{\sigma}{r}\Big)^{6}\,\Big],
\end{equation}
where $\varepsilon$ and $\sigma$ are the energy and length scales of
the fluid phase and $r_c\!=\!2.5\,\sigma$ is a cutoff radius. The
interaction between the wall atoms and fluid monomers is also
modeled by the LJ potential with parameters $\varepsilon_{\rm wf}$
(listed in Table\,\ref{tabela}) and $\sigma_{\rm wf}\,{=}\,\sigma$.
The wall atoms do not interact with each other via the LJ potential.

\begin{figure}[t]
\vspace*{-3mm}
\includegraphics[width=12.0cm,angle=0]{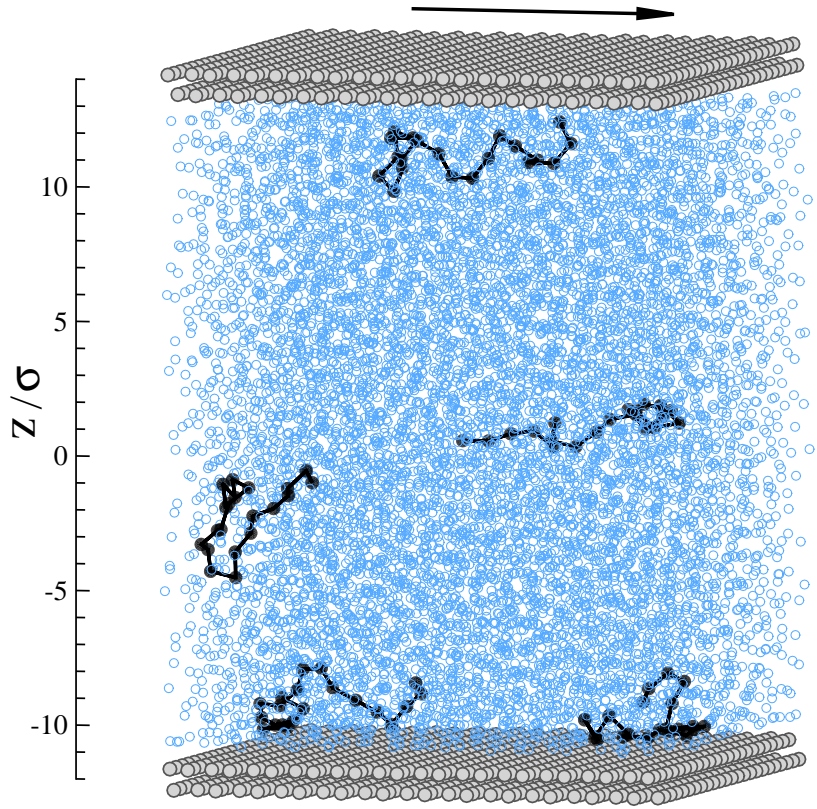}
\caption{(Color online) A snapshot of fluid monomers (open blue
circles) and wall atoms (filled gray circles) positions. Five
polymer chains are marked by solid lines and filled black circles.
The black arrow indicates the direction of the upper wall velocity
$U\,{=}\,\,0.5\,\sigma/\tau$. The fluid monomer density is
$\rho\,{=}\,0.89\,\sigma^{-3}$ and the wall density is
$\rho_{w}\,{=}\,1.80\,\sigma^{-3}$. The rest of parameters for the
system $5$ are given in Table\,\ref{tabela}.} \label{snapshot}
\end{figure}

Three types of fluid were considered in the present study, i.e.,
monomeric (or simple) fluid and polymer melts with the number of
monomers per chain $N\,{=}\,10$ and $N\,{=}\,20$. In the case of
polymers, the nearest-neighbor monomers in a chain interact through
the finitely extensible nonlinear elastic (FENE)
potential~\cite{Bird87}
\begin{equation}
V_{FENE}(r)=-\frac{k_s}{2}\,r_{\!o}^2\ln[1-r^2/r_{\!o}^2],
\end{equation}
with the energy and length parameters
$k_s\,{=}\,30\,\varepsilon\sigma^{-2}$ and
$r_{\!o}\,{=}\,1.5\,\sigma$ introduced by Kremer and
Grest~\cite{Kremer90}. Figure\,\ref{snapshot} shows a snapshot of an
unentangled polymer melt with linear flexible chains $N\,{=}\,20$
confined between solid walls.

The heat exchange between the fluid phase and the external heat bath
was regulated via a Langevin thermostat~\cite{Grest86}, which was
applied only to the direction of motion perpendicular to the plane
of shear~\cite{Thompson90}. The equations of motion for fluid
monomers in all three directions are given as follows:
\begin{eqnarray}
\label{Langevin_x}
m\ddot{x}_i & = & -\sum_{i \neq j} \frac{\partial V_{ij}}{\partial x_i}\,, \\
\label{Langevin_y}
m\ddot{y}_i + m\Gamma\dot{y}_i & = & -\sum_{i \neq j} \frac{\partial V_{ij}}{\partial y_i} + f_i\,, \\
\label{Langevin_z}
m\ddot{z}_i & = & -\sum_{i \neq j} \frac{\partial V_{ij}}{\partial z_i}\,, %
\end{eqnarray}
where the summation is performed over the fluid monomers and wall
atoms within the cutoff radius $r_c\!=\!2.5\,\sigma$,
$\Gamma\,{=}\,1.0\,\tau^{-1}$ is the friction coefficient, and $f_i$
is a random force with zero mean and variance $\langle
f_i(0)f_j(t)\rangle\,{=}\,\,2mk_BT\Gamma\delta(t)\delta_{ij}$
determined from the fluctuation-dissipation theorem. The thermostat
temperature is $T\,{=}\,1.1\,\varepsilon/k_B$, where $k_B$ is the
Boltzmann constant. The equations of motion were integrated using
the fifth-order gear-predictor algorithm~\cite{Allen87} with a time
step $\triangle t\,{=}\,0.002\,\tau$, where
$\tau\!=\!\sqrt{m\sigma^2/\varepsilon}$ is the characteristic time
of the LJ potential. The small time step $\triangle
t\,{=}\,0.002\,\tau$ was used in our previous
studies~\cite{Priezjev07,Priezjev08,Priezjev09} for similar MD
setups in order to compute accurately the trajectories of fluid
molecules and wall atoms near interfaces.

Each confining wall is composed of $1152$ atoms arranged in two
layers of the face-centered cubic (fcc) or body-centered cubic (bcc)
lattice. The wall density, lattice type, its orientation with
respect to the shear flow direction, and wall-fluid interaction
energy are reported in Table\,\ref{tabela}. The wall atoms were
either fixed at the lattice sites or were allowed to oscillate about
their equilibrium lattice positions under the harmonic potential
$V_{sp}\,{=}\,\frac{1}{2}\,\kappa\,r^2$ with the spring stiffness
coefficient $\kappa\,{=}\,1200\,\varepsilon/\sigma^2$. In the latter
case, the Langevin thermostat was applied to the $\hat{x}$,
$\hat{y}$ and $\hat{z}$ components of the wall atom equations of
motion. For example, the $\hat{x}$ component of the equation of
motion is given by
\begin{eqnarray}
\label{Langevin_wall_x} m_w\,\ddot{x}_i + m_w\,\Gamma\dot{x}_i & = &
-\sum_{i \neq j} \frac{\partial V_{ij}}{\partial x_i} -
\frac{\partial V_{sp}}{\partial x_i} + f_i\,,
\end{eqnarray}
where $m_w\,{=}\,10\,m$, the friction coefficient is
$\Gamma\,{=}\,1.0\,\tau^{-1}$ and the sum is taken over the
neighboring fluid monomers within the cutoff radius
$r_c\!\,\,{=}\,\,2.5\,\sigma$. Periodic boundary conditions were
imposed along the the $\hat{x}$ and $\hat{y}$ directions parallel to
the confining walls.

Initially, the fluid was equilibrated at a constant normal pressure
(shown in Table\,\ref{tabela}) for about $5\times10^4\tau$. Then,
the channel height was fixed and the system was additionally
equilibrated for $5\times10^4\tau$ at a constant density ensemble.
The steady flow was generated by moving the upper wall with a
constant speed $U$ in the $\hat{x}$ direction parallel to the
immobile lower wall (see Fig.\,\ref{schematic}). The lowest speed of
the upper wall is $U\,{=}\,\,0.05\,\sigma/\tau$. Both fluid velocity
and density profiles were computed within horizontal bins of
thickness $\Delta z\,{=}\,0.01\,\sigma$ for a time period up to
$6\times10^5\tau$.

\begin{table}[b]
\caption{The fluid monomer density $\rho$, number of monomers per
chain $N$, distance between the wall lattice planes in contact with
fluid $h$, wall area in the $xy$ plane, fluid pressure at
equilibrium (i.e., $U=0$), wall density $\rho_{w}$, lattice type,
Miller indices for the $xy$ plane, lattice orientation along the
shear flow direction ($\hat{x}$ direction), the $\hat{x}$ and
$\hat{y}$ components of the first reciprocal lattice vector
$\mathbf{G}_1(k_x,k_y)$, wall-fluid interaction energy, and the
spring stiffness coefficient for thermal walls.}
 \vspace*{3mm}
 \begin{ruledtabular}
 \begin{tabular}{r r r r r r r r r r r r r}
   $\#$ &  $\rho\,\sigma^3$ & $N$ & $h/\sigma$ & $A_{xy}/\sigma^2$ & $P/\varepsilon\sigma^{-3}$ & $\rho_{w}\sigma^3$ & ${\rm type}$ & $(ijk)$ & $\hat{x}\,~~$ & $(k_x\sigma,k_y\sigma)$ & $\varepsilon_{\rm wf}/\varepsilon$ & $\kappa/\varepsilon\sigma^{-2}$
   \\ [3pt] \hline 
   $1$ &      $0.91$ &  $20$ & $22.02$ & $502.28$ & $1.0~~$ & $1.40$  &  ${\rm fcc}$ & $(111)$  &  $[11\bar{2}]$ & $(7.23,0)$ & $0.9$ & ${\rm fixed}~$
   \\ [3pt]
   $2$ &      $0.91$ &  $20$ & $22.02$ & $502.28$ & $1.0~~$ & $1.40$  &  ${\rm fcc}$ & $(111)$  &  $[1\bar{1}0]$ & $(6.26,3.62)$ & $0.9$ & ${\rm fixed}~$
   \\ [3pt]
   $3$ &      $0.88$ &  $20$ & $19.46$ & $589.79$ & $0.5~~$ & $1.10$  &  ${\rm fcc}$ & $(111)$  &  $[11\bar{2}]$ & $(6.67,0)$ &  $0.8$ & $1200~$
   \\ [3pt]
   $4$ &      $0.88$ &  $20$ & $19.46$ & $589.79$ & $0.5~~$ & $1.10$  &  ${\rm fcc}$ & $(111)$  &  $[1\bar{1}0]$ & $(5.78,3.34)$ &  $0.8$ & $1200~$
   \\ [3pt]
   $5$ &      $0.89$ &  $20$ & $26.44$ & $424.73$ & $0.5~~$ & $1.80$  &  ${\rm fcc}$ & $(111)$  &  $[11\bar{2}]$ & $(7.86,0)$ &  $1.0$ & ${\rm fixed}~$
   \\ [3pt]
   $6$ &      $0.89$ &  $20$ & $26.44$ & $424.73$ & $0.5~~$ & $1.80$  &  ${\rm fcc}$ & $(111)$  &  $[1\bar{1}0]$ & $(6.81,3.93)$ &  $1.0$ & ${\rm fixed}~$
   \\ [3pt]
   $7$ &      $0.83$ &  $10$ & $23.93$ & $502.28$ & $0.0~~$ & $1.40$  &  ${\rm fcc}$ & $(111)$  &  $[11\bar{2}]$ & $(7.23,0)$ & $0.7$ & $1200~$
   \\ [3pt]
   $8$ &      $0.83$ &  $10$ & $23.93$ & $502.28$ & $0.0~~$ & $1.40$  &  ${\rm fcc}$ & $(111)$  &  $[1\bar{1}0]$ & $(6.26,3.62)$ & $0.7$ & $1200~$
   \\ [3pt]
   $9$ &      $0.88$ &  $20$ & $24.72$ & $459.42$ & $0.5~~$ & $1.60$  &  ${\rm fcc}$ &  $(111)$  &  $[11\bar{2}]$ & $(7.56,0)$ & $0.8$ & $1200~$
   \\ [3pt]
   $10$ &     $0.88$ &  $20$ & $24.72$ & $459.42$ & $0.5~~$ & $1.60$  &  ${\rm fcc}$ &  $(111)$  &  $[1\bar{1}0]$ & $(6.55,3.78)$ & $0.8$ & $1200~$
   \\ [3pt]
   $11$ &     $0.89$ &  $20$ & $19.12$ & $595.87$ & $0.5~~$ & $1.90$  &  ${\rm bcc}$ &  $(001)$  &  $[100]$ & $(6.18,0)$ &  $0.4$ & ${\rm fixed}~$
   \\ [3pt]
   $12$ &     $0.89$ &  $20$ & $19.12$ & $595.87$ & $0.5~~$ & $1.90$  &  ${\rm bcc}$ &  $(001)$  &  $[100]$ & $(6.18,0)$ &  $0.5$ & ${\rm fixed}~$
   \\ [3pt]
   $13$ &     $0.89$ &  $20$ & $19.12$ & $595.87$ & $0.5~~$ & $1.90$  &  ${\rm bcc}$ &  $(001)$  &  $[100]$ & $(6.18,0)$ &  $0.6$ & ${\rm fixed}~$
   \\ [3pt]
   $14$ &     $0.85$ &  $10$ & $19.98$ & $595.87$ & $0.5~~$ & $1.90$  &  ${\rm bcc}$ &  $(001)$  &  $[100]$ & $(6.18,0)$ &  $0.4$ & $1200~$
   \\ [3pt]
   $15$ &     $0.85$ &  $10$ & $19.98$ & $595.87$ & $0.5~~$ & $1.90$  &  ${\rm bcc}$ &  $(001)$  &  $[100]$ & $(6.18,0)$ &  $0.5$ & $1200~$
   \\ [3pt]
   $16$ &     $0.85$ &  $10$ & $19.98$ & $595.87$ & $0.5~~$ & $1.90$  &  ${\rm bcc}$ &  $(001)$  &  $[100]$ & $(6.18,0)$ &  $0.6$ & $1200~$
   \\ [3pt]
   $17$ &     $0.81$ &  $1$  & $34.86$ & $350.61$ & $2.36~~$ & $2.40$  &  ${\rm fcc}$ & $(111)$  &  $[11\bar{2}]$ & $(8.65,0)$ & $0.4$ & ${\rm fixed}~$
   \\ [3pt]
   $18$ &     $0.81$ &  $1$  & $34.86$ & $350.61$ & $2.36~~$ & $2.40$  &  ${\rm fcc}$ & $(111)$  &  $[1\bar{1}0]$ & $(7.49,4.33)$ & $0.4$ & ${\rm fixed}~$
   \\ [3pt]
   $19$ &     $0.81$ &  $1$  & $34.86$ & $350.61$ & $2.36~~$ & $2.40$  &  ${\rm fcc}$ & $(111)$  &  $[11\bar{2}]$ & $(8.65,0)$ & $0.3$ & ${\rm fixed}~$
   \\ [3pt]
   $20$ &     $0.81$ &  $1$  & $34.86$ & $350.61$ & $2.36~~$ & $2.40$  &  ${\rm fcc}$ & $(111)$  &  $[1\bar{1}0]$ & $(7.49,4.33)$ & $0.3$ & ${\rm fixed}~$
   \\ [3pt]
 \end{tabular}
 \end{ruledtabular}
 \label{tabela}
\end{table}

\section{Results}
\label{sec:Results}

\subsection{Fluid density and velocity profiles}

\begin{figure}[t]
\includegraphics[width=12.cm,angle=0]{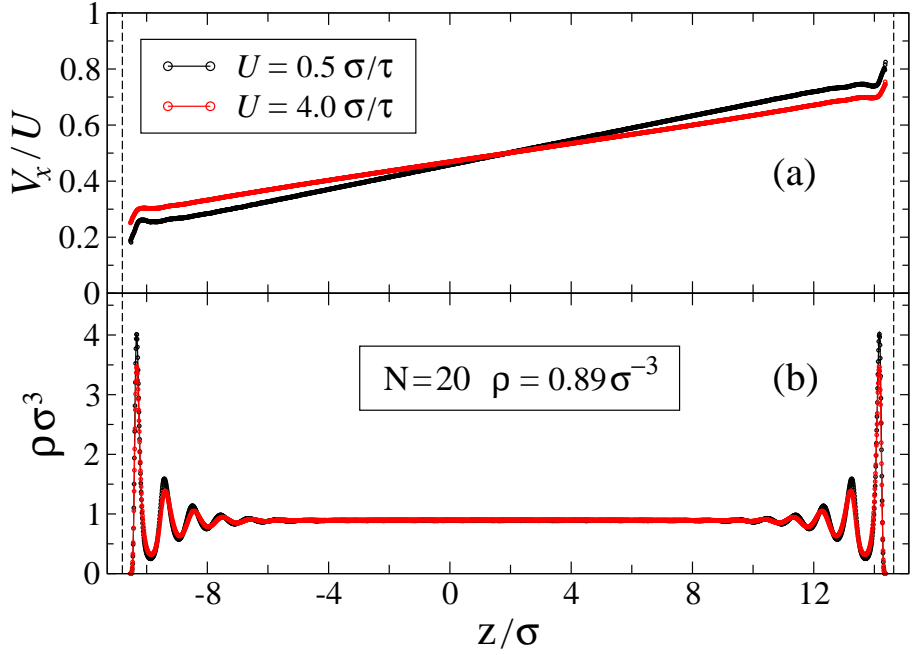}
\caption{(Color online) Averaged normalized velocity (a) and density
(b) profiles across the channel for the upper wall speeds
$U\,{=}\,\,0.5\,\sigma/\tau$ and $U\,{=}\,\,4.0\,\sigma/\tau$. The
uniform monomer density of the polymer melt $N\,{=}\,20$ away from
the walls is $\rho\,{=}\,0.89\,\sigma^{-3}$ (system 5 in
Table\,\ref{tabela}). The vertical axes indicate the location of the
fcc lattice planes (at $z/\sigma\,\,{=}\,-11.30$ and $15.14$) in
contact with the fluid. The dashed lines at
$z/\sigma\,\,{=}\,-10.80$ and $14.64$ denote reference planes for
computing the slip length.} \label{velo_dens}
\end{figure}

The averaged fluid density and velocity profiles are presented in
Fig.\,\ref{velo_dens} for the upper wall speeds
$U\,{=}\,\,0.5\,\sigma/\tau$ and $U\,{=}\,\,4.0\,\sigma/\tau$. The
density profiles exhibit a typical layered structure which extends
for about $5\,\sigma-6\,\sigma$ away from the solid walls. The
amplitude of the first peak in the density profile determines the
contact density $\rho_c$. We emphasize that the thickness of the
averaging bins $\Delta z\,{=}\,0.01\,\sigma$ is small enough so that
the magnitude of the peaks does not depend on the bin thickness and
bin location relative to the walls. On the other hand, the shape of
the density profiles will remain unchanged if thinner bins are used;
however, the averaging is computationally more expensive. As evident
from Fig.\,\ref{velo_dens}\,(b), the contact density is reduced at
higher upper wall speeds.

Two representative velocity profiles normalized by the upper wall
speed are shown in Fig.\,\ref{velo_dens}\,(a). The slip velocity
increases at higher upper wall speeds. The location of the
liquid-solid interface (marked by the dashed vertical lines in
Fig.\,\ref{velo_dens}) is defined at the distance $0.5\,\sigma$ away
from the wall lattice planes to take into account the excluded
volume due to wall atoms. The slip length was computed from the
linear fit to the velocity profiles excluding regions of about
$2\,\sigma$ from the solid walls. At all shear rates examined in the
present study, the velocity profiles are linear across the channel
and the slip length is larger than about $3\,\sigma$.

\subsection{Rate dependence of viscosity and slip length}

\begin{figure}[t]
\includegraphics[width=12.cm,angle=0]{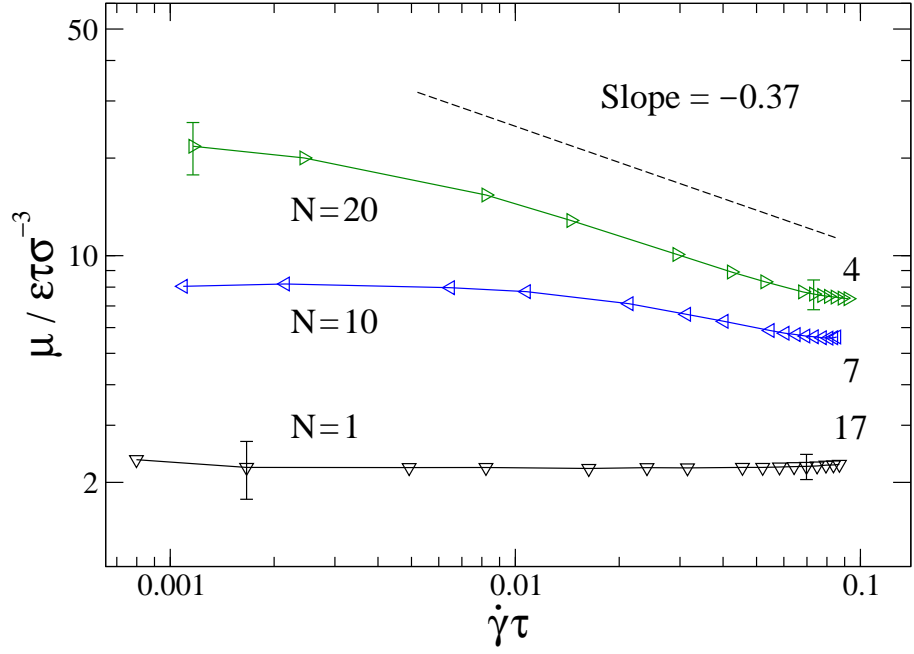}
\caption{(Color online) Shear rate dependence of the fluid viscosity
$\mu$ (in units of $\varepsilon\tau\sigma^{-3}$) for the indicated
systems (listed in Table\,\ref{tabela}). The dashed line with a
slope $-0.37$ is shown for reference. Solid curves are a guide for
the eye.} \label{visc_shear_all}
\end{figure}

The fluid viscosity was estimated from the relation between shear
rate and shear stress which was computed using the Kirkwood
formula~\cite{Kirkwood}. The variation of viscosity as a function of
shear rate is presented in Fig.\,\ref{visc_shear_all} for selected
systems listed in Table\,\ref{tabela}. In agreement with previous
studies with a similar setup~\cite{Nature97,Priezjev07,PriezjevJCP},
the viscosity of monatomic fluids is independent of shear rate and
equals $\mu\,{=}\,(2.2\pm0.2)\,\,\varepsilon\tau\sigma^{-3}$ when
the fluid density is $\rho\,{=}\,0.81\,\sigma^{-3}$. As expected,
the shear viscosity of polymer melts with chains $N\,{=}\,10$ and
$N\,{=}\,20$ is higher than the viscosity of simple monatomic
fluids. For similar flow conditions, the transition from a Newtonian
to a shear-thinning flow regime occurs at lower shear rates for
polymers with longer chains $N\,{=}\,20$ because of their slower
intrinsic relaxation. The slope of the shear-thinning region $-0.37$
shown in Fig.\,\ref{visc_shear_all} is consistent with the results
reported in earlier studies for polymer melts $N\,{=}\,20$ at
different densities~\cite{Priezjev08,Priezjev09}. The errors arising
from averaging over thermal fluctuations are greater at lower shear
rates.

\begin{figure}[t]
\includegraphics[width=12.cm,angle=0]{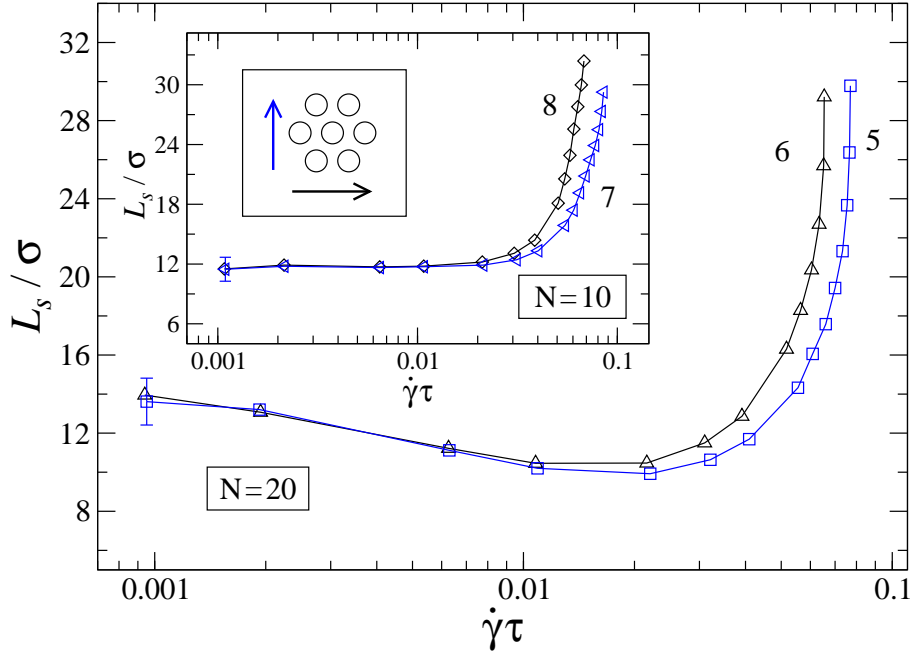}
\caption{(Color online) Slip length $L_s/\sigma$ as a function of
shear rate for polymer melts with chains $N\,{=}\,20$ and
$N\,{=}\,10$ (see inset). The system parameters are listed in
Table\,\ref{tabela}. Open circles in the inset represent the $(111)$
face of the fcc lattice atoms in contact with the fluid. The
vertical blue arrow indicates the shear flow direction with respect
to the $[11\bar{2}]$ fcc lattice orientation (systems 5 and 7). The
horizontal black arrow shows the flow direction with respect to the
$[1\bar{1}0]$ orientation (systems 6 and 8).} \label{shear_ls_all}
\end{figure}

The nonlinear rate dependence of the slip length is shown in
Fig.\,\ref{shear_ls_all} for polymer melts with chains $N\,{=}\,10$
and $N\,{=}\,20$. The shear flow direction is oriented along the
crystallographic axis of the $(111)$ plane of the fcc wall lattice
for systems $6$ and $8$ (see Table\,\ref{tabela}). In contrast, the
fcc lattice plane is rotated by $90^{\circ}$ with respect to the
flow direction for systems $5$ and $7$ as indicated by open circles
and blue vertical arrow in the inset of Fig.\,\ref{shear_ls_all}. At
low shear rates $\dot{\gamma}\tau\lesssim0.02$, the slip length is
independent of the wall lattice orientation relative to the shear
flow direction; while at higher shear rates, the slip length is
greater when the shear flow is parallel to the crystallographic axis
of the triangular lattice. The same trend for the slip length is
observed for monatomic fluids (not shown). These results demonstrate
that at sufficiently high shear rates the slip flow is anisotropic
even for atomically flat crystalline surfaces.

The appearance of the local minimum in the rate dependence of the
slip length reported in Fig.\,\ref{shear_ls_all} for polymer melts
with chains $N\,{=}\,20$ was explained in the previous MD
study~\cite{Priezjev08}. The initial decay of the slip length at low
shear rates is associated with a slight decrease in the melt
viscosity while the friction coefficient at the liquid-solid
interface ($k\,{=}\,\,\mu/L_s$) remains constant (see also next
section). With increasing shear rate, the friction coefficient
decreases faster than the shear viscosity, and, therefore, the slip
length grows rapidly~\cite{Priezjev08,Niavarani08}. Since the
transition to the shear-thinning regime occurs at higher shear rates
for polymer melts with shorter chains $N\,{=}\,10$, the slip length
remains nearly constant at low shear rates and then increases
rapidly at higher rates (see inset in Fig.\,\ref{shear_ls_all}).
These results agree well with the previous simulation results for
slip flow of polymers with chain lengths $N\leqslant16$ and lower
fluid density~\cite{Priezjev04}.

\subsection{Friction coefficient versus slip velocity}

The relation between the slip length and shear rate can be expressed
in terms of the friction coefficient at the liquid-solid interface
and slip velocity. In steady-state shear flow, the shear stress in
the bulk of the film ($\dot{\gamma}\mu$) is equal to the wall shear
stress ($kV_s$). In addition, if the velocity profile is linear
across the channel, then by definition $V_s\,{=}\,\,\dot{\gamma}L_s$
and the friction coefficient is given by $k\,{=}\,\,\mu/L_s$. In the
previous MD study, the slip flow of polymer melts with chains
$N\,{=}\,20$ was studied in the range of fluid densities
$0.86\leqslant\rho\,\sigma^3\leqslant1.02$, and velocity profiles
were found to be linear at all shear rated
examined~\cite{Priezjev08}. Furthermore, the friction coefficient
($k\,{=}\,\,\mu/L_s$) as a function of the slip velocity could be
well fitted by the following equation:
\begin{equation}
k/k^{\ast}=[1+(V_s/V_s^{\ast})^2]^{-0.35}, \label{friction_law}
\end{equation}
where $k^{\ast}$ is the friction coefficient at small slip
velocities when $V_s\ll V_s^{\ast}$ and $V_s^{\ast}$ is the
characteristic slip velocity that determines the onset of the
nonlinear regime~\cite{Priezjev08}. In the later study, the
simulations were performed at higher melt densities and the velocity
profiles at low shear rates were curved near interfaces, and, as a
result, the definition $V_s\,{=}\,\,\dot{\gamma}L_s$ could not be
applied~\cite{Priezjev09}. Therefore, the friction coefficient was
computed directly from the ratio of the wall shear stress and slip
velocity of the first fluid layer. For polymer melt densities
$\rho\,{=}\,1.04\,\sigma^{-3}$ and $1.06\,\sigma^{-3}$, the data
were also well described by Eq.\,(\ref{friction_law}), while at
higher melt densities ($1.08\leqslant\rho\,\sigma^3\leqslant1.11$)
only the nonlinear regime was observed~\cite{Priezjev09}.

In the present study, we extend the analysis of the friction
coefficient at the interface between crystalline walls and polymeric
fluids described by the parameters listed in Table\,\ref{tabela}.
Figure\,\ref{friction_velo} shows the friction coefficient as a
function of the slip velocity normalized by the parameters
$k^{\ast}$ and $V_s^{\ast}$ respectively. The data for all systems
in Table\,\ref{tabela} are well fitted by Eq.\,(\ref{friction_law})
over about three orders of magnitude. We also noticed the inverse
correlation between the friction coefficient $k^{\ast}$ and the
characteristic slip velocity $V_s^{\ast}$ (shown in
Fig.\,\ref{friction_velo_norm}). Note that for every two systems
with different orientation of the fcc wall lattice, the values of
$k^{\ast}$ are nearly the same, but the slip velocity $V_s^{\ast}$
is slightly smaller when the shear flow direction is parallel to the
crystallographic axis. This is consistent with the dynamic response
of the slip length reported in Fig.\,\ref{shear_ls_all} for two
different orientations of the fcc lattice.

\begin{figure}[t]
\includegraphics[width=12.cm,angle=0]{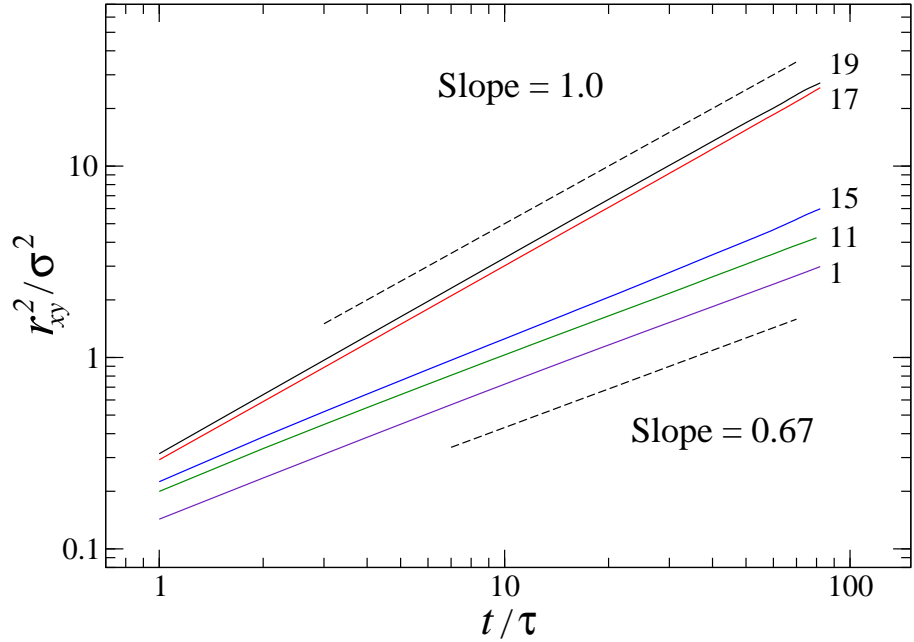}
\caption{(Color online) The mean square displacement of monomers in
the first fluid layer at equilibrium (i.e., $U=0$) as a function of
time $t$ (in units of $\tau$) for selected systems listed in
Table\,\ref{tabela}. The dashed lines are shown for reference.}
\label{msd}
\end{figure}

We next argue that the onset of the nonlinear regime in
Eq.\,(\ref{friction_law}) is determined by the intrinsic relaxation
time of the fluid monomers in the first layer near the solid wall.
In Figure\,\ref{msd} we plot the mean square displacement of fluid
monomers within the first layer for selected systems in
Table\,\ref{tabela} at equilibrium (i.e., when both walls are at
rest). The displacement as a function of time was computed along the
trajectory of a fluid monomer only if it remained in the first fluid
layer during the time interval between successive measurements of
the monomer position. For monatomic fluids, there is a linear
dependence between the mean square displacement and time, and,
consequently, the diffusion coefficient is well defined. This
behavior agrees well with the exponential relaxation of the
density-density autocorrelation function evaluated at the wavevector
of about $2\pi/\sigma$ in the first layer of monatomic fluids
confined by atomistic walls~\cite{Barrat99fd,Priezjev04}. In
contrast, fluid monomers that belong to a polymer chain diffuse
slower than monomers in simple fluids since their dynamics is
bounded by diffusion of the center of mass of the polymer chain. The
slope of the subdiffusive regime is shown in Fig.\,\ref{msd} by the
straight dashed line.

\begin{figure}[t]
\includegraphics[width=12.cm,angle=0]{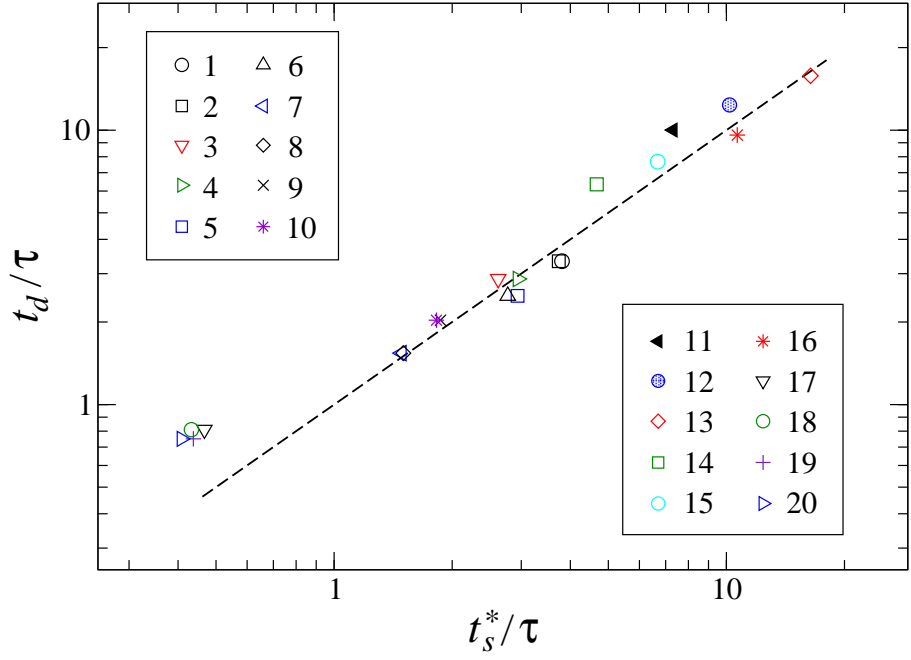}
\caption{(Color online) A correlation between the characteristic
slip time $t_s^{\ast}$ of the first fluid layer and the diffusion
time $t_d$ of fluid monomers between nearest minima of the surface
potential. The system parameters are given in Table\,\ref{tabela}.
The dashed line $y=x$ is shown as a reference. } \label{times}
\end{figure}

Finally, the comparison of the characteristic slip time of the first
fluid layer and the diffusion time of fluid monomers between nearest
minima of the surface potential is presented in Fig.\,\ref{times}.
The diffusion time was estimated from the mean square displacement
of fluid monomers in the first layer at the distance between nearest
minima of the periodic surface potential. The same distance divided
by the slip velocity $V_s^{\ast}$ defines the characteristic slip
time of the first fluid layer. In the case when the shear flow
direction is parallel to the $[1\bar{1}0]$ fcc lattice orientation
(e.g., see inset in Fig.\,\ref{shear_ls_all}), the slipping distance
of the first layer was computed by projecting the vector, which
connects nearest minima of the surface potential, onto the direction
of flow. Figure\,\ref{times} shows a strong correlation between the
characteristic slip time of the adjacent fluid layer and the
diffusion time of fluid monomers in that layer at equilibrium. These
results indicate that the linear-response regime in
Eq.\,(\ref{friction_law}) holds when the slip velocity is smaller
than the diffusion velocity of fluid monomers in contact with flat
crystalline walls.

\subsection{Friction coefficient and induced fluid structure}

The fluid structure near flat solid walls is characterized by
density layering perpendicular to the surface and ordering of fluid
monomers within the layers~\cite{Kaplan06}. Examples of oscillatory
density profiles in a polymer melt near confining walls were
presented in Fig.\,\ref{velo_dens}. It is intuitively expected that
enhanced fluid density layering normal to the surface (obtained, for
example, by increasing fluid pressure or wall-fluid interaction
energy) would correspond to a larger friction coefficient at the
liquid-solid interface. However, this correlation does not always
hold; for example, the amplitude of fluid density oscillations near
flat structureless walls might be large, but the friction
coefficient is zero. As emphasized in the original paper by Thompson
and Robbins~\cite{Thompson90}, the surface-induced fluid ordering
within the first layer of monomers correlates well with the degree
of slip at the liquid-solid interface. The measure of the induced
order in the adjacent fluid layer is the static structure factor,
which is defined as follows:
\begin{equation}
S(\mathbf{k})=\frac{1}{N_{\ell}}\,\,\Big|\sum_{j=1}^{N_{\ell}}
e^{i\,\mathbf{k}\cdot\mathbf{r}_j}\Big|^2,
\label{structure_factor}
\end{equation}
where $\mathbf{k}$ is a two-dimensional wavevector,
$\mathbf{r}_j\,{=}\,(x_j,y_j)$ is the position vector of the $j$-th
monomer, and $N_{\ell}$ is the number of monomers within the
layer~\cite{Thompson90}. The probability of finding fluid monomers
is greater near the minima of the periodic surface potential, and,
therefore, the structure factor typically contains a set of sharp
peaks at the reciprocal lattice vectors. It is well established that
the magnitude of the largest peak at the first reciprocal lattice
vector is one of the main factors that determine the value of the
slip length at the interface between flat crystalline surfaces and
monatomic fluids~\cite{Thompson90,Barrat99fd,Priezjev07,PriezjevJCP}
or polymer melts~\cite{Thompson95,Priezjev04,Priezjev08,Priezjev09}.

\begin{figure}[t]
\includegraphics[width=12.cm,angle=0]{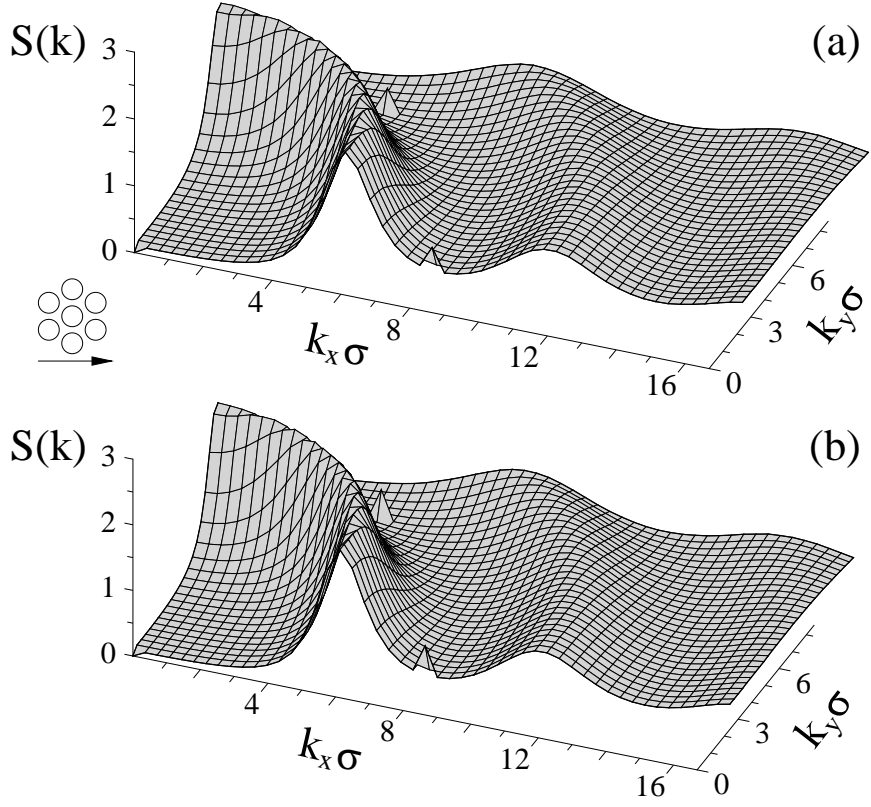}
\caption{Two-dimensional structure factor $S(k_x,k_y)$ computed in
the first fluid layer for $N\,{=}\,1$ and
$U\,{=}\,\,0.05\,\sigma/\tau$ [\,systems (a)\,$19$ and (b)\,$17$ in
Table\,\ref{tabela}\,]. The wall-fluid interaction energy is (a)
$\varepsilon_{\rm wf}\,{=}\,0.3\,\varepsilon$ and (b)
$\varepsilon_{\rm wf}\,{=}\,0.4\,\varepsilon$. The shear flow
direction (denoted by the horizontal arrow) is parallel to the
$[11\bar{2}]$ orientation of the $(111)$ face of the fcc wall
lattice (open circles).} \label{sk_simple}
\end{figure}

Next, we discuss the influence of the wall-fluid interaction energy,
wall lattice type and orientation, and slip velocity on the
structure factor computed in the first fluid layer. The effect of
the wall-fluid interaction energy is illustrated in
Fig.\,\ref{sk_simple} for monatomic fluids in contact with the
$(111)$ plane of the fcc wall lattice. The height of the
surface-induced peaks in the structure factor is slightly larger at
higher surface energy. The magnitude of the peak in the shear flow
direction is $S(8.65,0)\,{=}\,0.98$ for $\varepsilon_{\rm
wf}\,{=}\,0.3\,\varepsilon$ and $S(8.65,0)\,{=}\,1.06$ for
$\varepsilon_{\rm wf}\,{=}\,0.4\,\varepsilon$. Notice that the
height of the circular ridge characteristic of short range ordering
of fluid monomers is larger than the amplitude of the induced peaks
at the reciprocal lattice vectors. A similar trend in the height of
the peaks in the structure factor was observed previously for
monatomic fluids confined by the fcc walls with higher density
$\rho_{w}\,{=}\,2.73\,\sigma^{-3}$~\cite{Priezjev07}.

\begin{figure}[t]
\includegraphics[width=12.cm,angle=0]{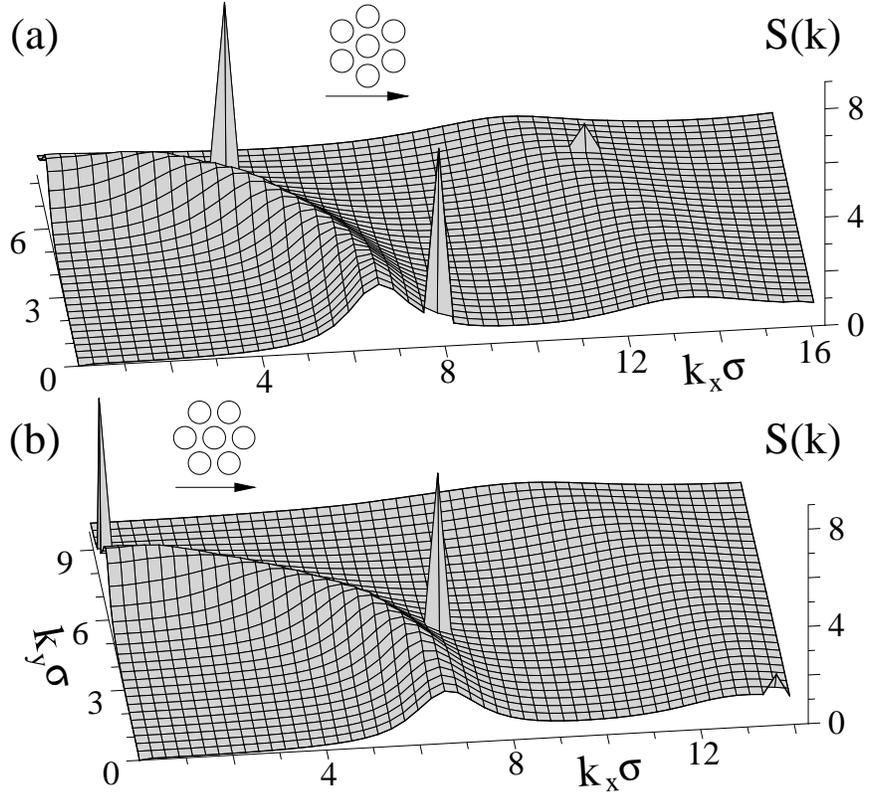}
\caption{Structure factor $S(k_x,k_y)$ averaged in the first fluid
layer for $N\,{=}\,20$ polymer systems (a)\,$5$ and (b)\,$6$ (see
parameters in Table\,\ref{tabela}). In each case, horizontal arrows
indicate the shear flow direction with respect to the orientation of
the $(111)$ plane of the fcc wall lattice (denoted by open circles).
The upper wall speed is $U\,{=}\,\,0.05\,\sigma/\tau$ in both
cases.} \label{sk_n20}
\end{figure}

\begin{figure}[t]
\includegraphics[width=12.cm,angle=0]{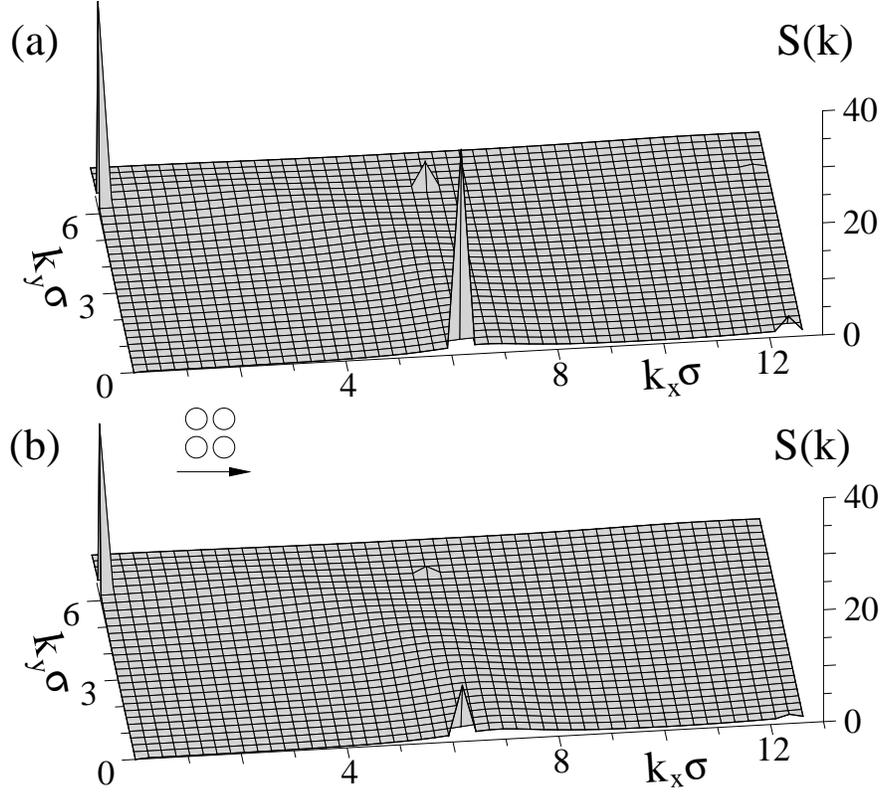}
\caption{Structure factor $S(k_x,k_y)$ computed in the first fluid
layer for $N\,{=}\,20$ polymer system $12$ (see
Table\,\ref{tabela}). The upper wall speed and slip velocity are (a)
$U\,{=}\,\,0.05\,\sigma/\tau$ and $V_s\,{=}\,\,0.012\,\sigma/\tau$
and (b) $U\,{=}\,\,2.0\,\sigma/\tau$ and
$V_s\,{=}\,\,0.51\,\sigma/\tau$, respectively. The horizontal arrows
denote the shear flow direction with respect to the orientation of
the $(001)$ face of the bcc wall lattice (open circles).}
\label{sk_n20bcc}
\end{figure}

Figure\,\ref{sk_n20} shows the structure factor computed in the
first fluid layer for polymer melts with chains $N\,{=}\,20$ in
contact with the $(111)$ plane of the fcc wall lattice. Due to the
hexagonal symmetry of the lattice, the structure factor exhibits six
peaks at the shortest reciprocal lattice vectors. Note that only two
main peaks are present in the first quadrant. The magnitude of the
peaks is the same at small slip velocities. The lattice orientation
with respect to the shear flow direction determines the location of
the main peaks. Finally, the effect of slip velocity on the
magnitude of the substrate-induced peaks in the structure factor is
presented in Fig.\,\ref{sk_n20bcc} for the polymer melt near the
$(001)$ plane of the bcc wall lattice. With increasing slip
velocity, the height of the induced peak along the shear flow
direction decreases significantly, whereas the magnitude of the peak
in the perpendicular direction is less affected by slip. In contrast
to monatomic fluids, the amplitude of the peak due to short range
order of monomers that belong to polymer chains is much smaller than
the magnitude of the induced peaks at the shortest reciprocal
lattice vectors (see Fig.\,\ref{sk_n20bcc}).

The correlation between surface-induced structure in the first fluid
layer and the friction coefficient was investigated previously for
polymer melts with chains $N\,{=}\,20$ confined by atomically flat
walls~\cite{Priezjev08,Priezjev09}. The simulations were performed
at fluid densities $0.86\leqslant\rho\,\sigma^3\leqslant1.11$ and
the wall density $\rho_{w}\,{=}\,1.40\,\sigma^{-3}$. It was found
that the data for the friction coefficient at different shear rates
and fluid densities collapsed onto a master curve when plotted as a
function of a variable $S(0)/[S(\mathbf{G}_1)\,\rho_c]$, where
$\mathbf{G}_1$ is the first reciprocal lattice vector in the shear
flow direction~\cite{Priezjev08,Priezjev09}. The collapse of the
data holds at relatively small values of the friction coefficient
$k\lesssim4\,\varepsilon\tau\sigma^{-4}$ and for slip lengths larger
than approximately $5\,\sigma$. Although these results are
promising, the simulations were limited to a single wall density and
the $[11\bar{2}]$ orientation of the $(111)$ plane of the fcc wall
lattice.

\begin{figure}[t]
\includegraphics[width=12.cm,angle=0]{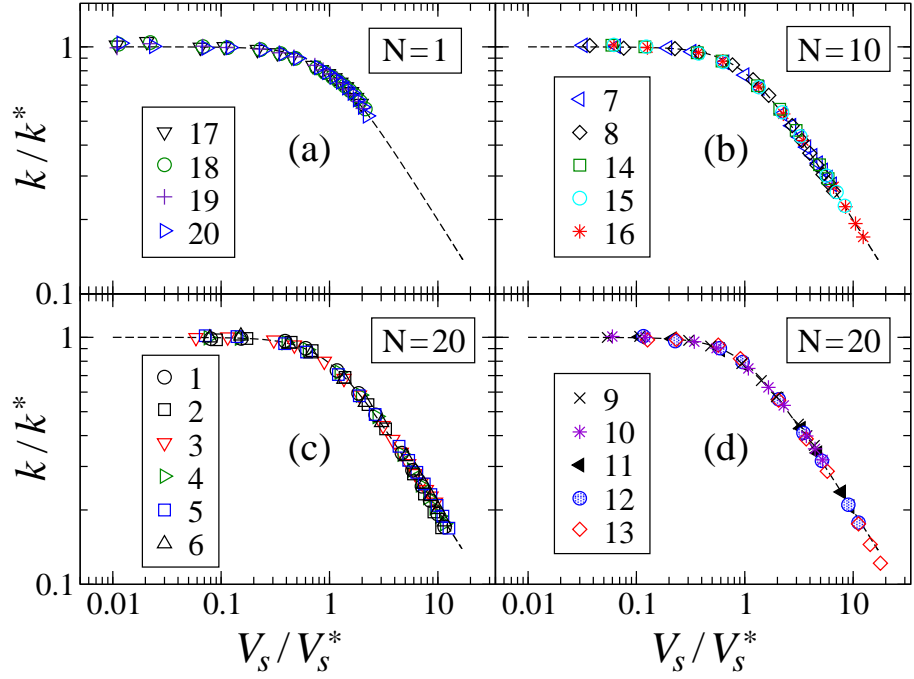}
\caption{(Color online) Log-log plot of the friction coefficient
$k\,{=}\,\,\mu/L_s$ (in units of $\varepsilon\tau\sigma^{-4}$) as a
function of the slip velocity $V_s=L_s\dot{\gamma}$ (in units of
$\sigma/\tau$) for systems listed in Table\,\ref{tabela}. The values
of the normalization parameters $V_s^{\ast}$ and $k^{\ast}$ are
presented in Fig.\,\ref{friction_velo_norm}. The dashed curves
$y=(1+x^2)^{-0.35}$ are the best fit to the data.}
\label{friction_velo}
\end{figure}

\begin{figure}[t]
\includegraphics[width=12.cm,angle=0]{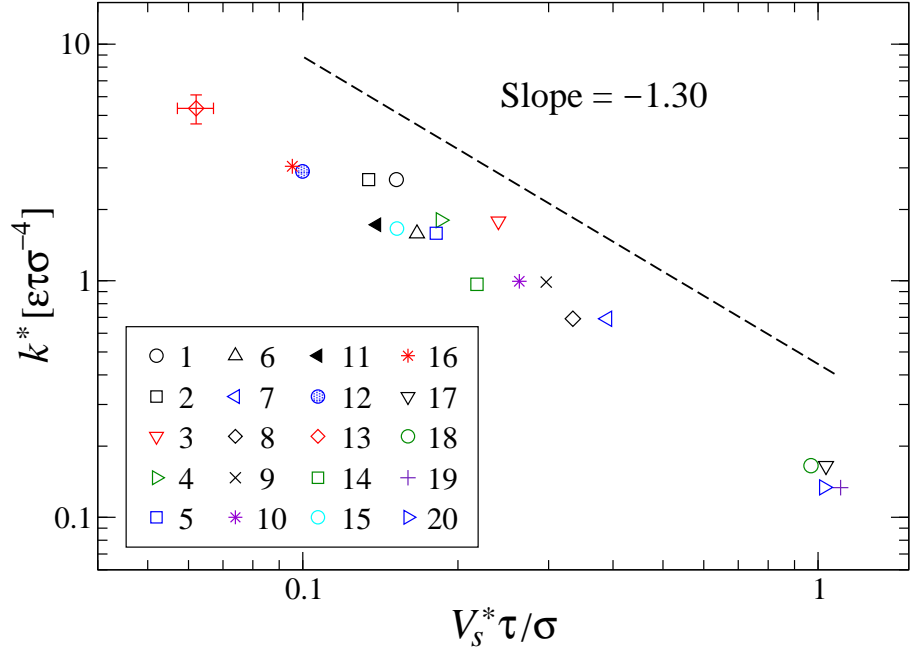}
\caption{(Color online) The normalization parameters $V_s^{\ast}$
(in units of $\sigma/\tau$) and $k^{\ast}$ (in units of
$\varepsilon\tau\sigma^{-4}$) used to fit the data in
Fig.\,\ref{friction_velo} to Eq.\,(\ref{friction_law}). The indices
in the inset denote systems listed in Table\,\ref{tabela}. The
dashed line with a slope $-1.30$ is shown for reference.}
\label{friction_velo_norm}
\end{figure}

\begin{figure}[t]
\includegraphics[width=12.cm,angle=0]{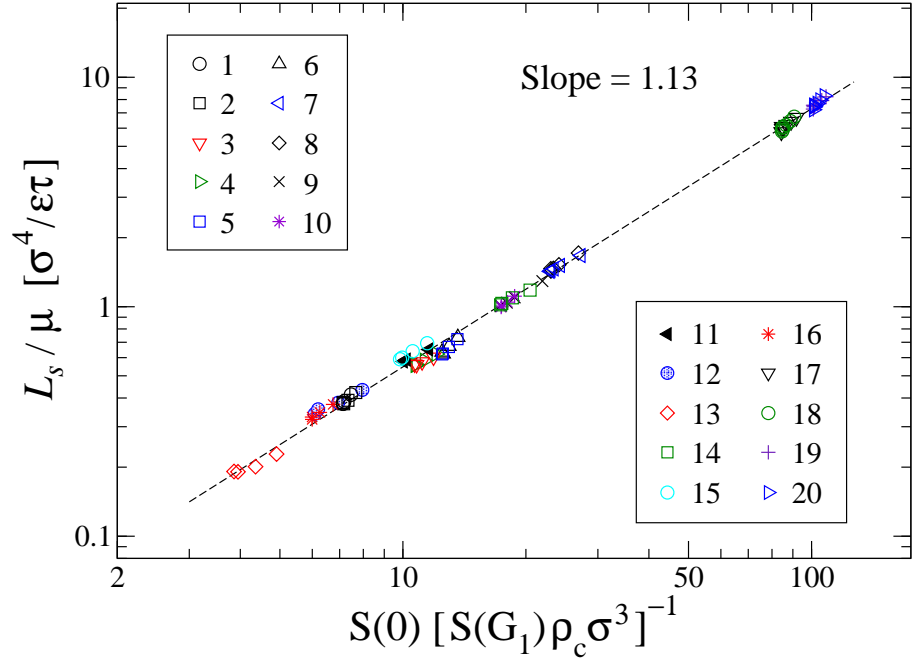}
\caption{(Color online) Log-log plot of the ratio $L_s/\mu$ (in
units of $\sigma^4/\varepsilon\tau$) as a function of the variable
$S(0)/\,[S(\mathbf{G}_1)\,\rho_c]$ computed in the first fluid layer
at low shear rates. The system parameters are listed in
Table\,\ref{tabela}. The dashed line $y=0.041\,x^{1.13}$ is the best
fit to the data.} \label{inv_fr_vs_S0_div_S7_ro_c_low}
\end{figure}

In the present study, a number of parameters that affect slippage at
the liquid-solid interface have been examined, i.e., fluid and wall
densities, polymer chain length, wall lattice type and orientation,
wall-fluid interaction energy, thermal and solid walls (see
Table\,\ref{tabela}). We first consider the linear-response regime
where the friction coefficient weakly depends on the slip velocity
($k/k^{\ast}\gtrsim 0.8$ in Fig.\,\ref{friction_velo}).
Figure\,\ref{inv_fr_vs_S0_div_S7_ro_c_low} shows the ratio $L_s/\mu$
(an inverse friction coefficient) as a function of the variable
$S(0)/[S(\mathbf{G}_1)\,\rho_c]$ computed in the first fluid layer
for twenty systems listed in Table\,\ref{tabela}. In the case when
the shear flow direction is parallel to the $[1\bar{1}0]$
orientation of the fcc lattice [e.g., see Fig.\,\ref{sk_n20}\,(b)],
the structure factor was computed at the shortest reciprocal lattice
vector $\mathbf{G}_1$ aligned at an angle of $30^{\circ}$ with
respect to the $\hat{x}$ axis. Since the magnitude of the
surface-induced peaks in the structure factor scales with the number
of monomers in the first fluid layer, the height of the main peak
$S(\mathbf{G}_1)$ was normalized by the average number of monomers
in the layer $N_{\ell}\,{=}\,S(0)$. The data in
Fig.\,\ref{friction_velo} are well described by a power-law fit with
the slope $1.13$. These results suggest that at the interface
between simple or polymeric fluids and flat crystalline surfaces,
the ratio of the slip length and viscosity at low shear rates [\,or
the value of parameter $k^{\ast}$ in Eq.\,(\ref{friction_law})\,]
can be estimated from equilibrium measurements of the structure
factor and the contact density of the first fluid layer.

\begin{figure}[t]
\includegraphics[width=12.cm,angle=0]{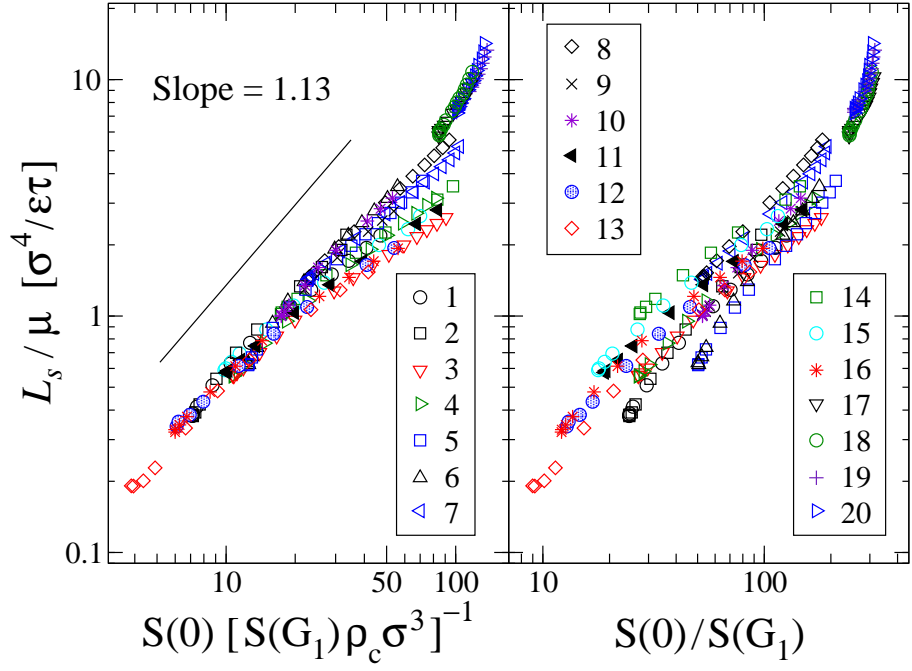}
\caption{(Color online) Log-log plot of the ratio $L_s/\mu$ (in
units of $\sigma^4/\varepsilon\tau$) as a function of variables (a)
$S(0)/\,[S(\mathbf{G}_1)\,\rho_c]$ and (b) $S(0)/S(\mathbf{G}_1)$
computed in the first fluid layer at all shear rates examined. The
system parameters are given in Table\,\ref{tabela}. The black line
with a slope $1.13$ is shown for reference.}
\label{friction_structure_full}
\end{figure}

In Figure\,\ref{friction_structure_full} we report the dependence of
the friction coefficient ($k\,{=}\,\,\mu/L_s$) on the structure
factor and contact density of the first fluid layer at all shear
rates examined in this study (the same data as in
Fig.\,\ref{friction_velo}). Note that at higher shear rates the
derivative of $L_s/\mu$ with respect to
$S(0)/[S(\mathbf{G}_1)\,\rho_c]$ for several systems listed in
Table\,\ref{tabela} deviates significantly from the slope $1.13$
shown for reference in Fig.\,\ref{friction_structure_full}\,(a). In
addition, for any two systems with the same $\rho_{w}$ and
$\varepsilon_{\rm wf}$, the ratio $L_s/\mu$ as a function of
$S(0)/[S(\mathbf{G}_1)\,\rho_c]$ depends on the orientation of the
fcc wall lattice with respect to the shear flow direction. Although
the data in Fig.\,\ref{friction_structure_full}\,(b) are somewhat
scattered, the results show the same trend, namely, the friction
coefficient decreases when the magnitude of the normalized peak in
the structure factor is reduced. The collapse of the data for $L_s$
versus $S(\mathbf{G}_1)/S(0)$ was reported in
Ref.\,\cite{Thompson90} for monatomic fluids and crystalline walls
when $L_s\lesssim 3.5\,\sigma$ and the boundary conditions are rate
independent. In the present study, the slip lengths are greater than
about $5\,\sigma$ except for the systems $13$ and $16$ where
$L_s\approx3\,\sigma$ at low shear rates. We finally comment that
our results were not analyzed with respect to the relation (between
the friction coefficient and induced fluid structure and in-plane
diffusion coefficient) derived in Ref.\,\cite{Barrat99fd} for simple
fluids, because the slope of the mean square displacement versus
time for polymer systems (shown in Fig.\,\ref{msd}) is less than one
and thus the diffusion coefficient is not well defined.

\section{Conclusions}

In this paper, we investigated the dynamic behavior of the slip
length at interfaces between polymeric or monatomic fluids and flat
crystalline surfaces using molecular dynamics simulations. The
polymer melt was modeled as a collection of bead-spring linear
flexible chains below the entanglement length. We considered shear
flow conditions at relatively low fluid densities (pressures) and
weak wall-fluid interaction energies so that fluid velocity profiles
are linear across the channel at all shear rates examined. It was
found that the slip length does not depend on the wall lattice
orientation with respect to the flow direction only at low shear
rates, whereas the slip is enhanced at high shear rates when the
flow direction is parallel to the crystallographic axis of the
substrate.

In the steady shear flow of either monatomic fluids or polymer
melts, the friction coefficient at the liquid-solid interface
(computed from the ratio of fluid viscosity and slip length)
undergoes a transition from a constant value to the power-law decay
as a function of the slip velocity. The characteristic velocity of
the transition is determined by the diffusion of fluid monomers over
the distance between nearest minima of the substrate potential. It
is demonstrated that the friction coefficient at small slip
velocities is a function of the magnitude of the surface-induced
peak in the structure factor and the contact density of the first
fluid layer. These conclusions hold for different wall and fluid
densities, chain lengths, surface energies, lattice types and
orientations, thermal or solid walls.

\section*{Acknowledgments}

Financial support from the National Science Foundation and the
Petroleum Research Fund of the American Chemical Society is
gratefully acknowledged. Computational work in support of this
research was performed at Michigan State University's High
Performance Computing Facility.

\bibliographystyle{prsty}

\end{document}